\documentstyle[prb,aps,preprint,tighten,eqsecnum]{revtex}

\begin{document}
\preprint{cond-mat/9501025}
\draft
\title{Generalized circular ensemble of scattering matrices \\ for a chaotic
cavity with non-ideal leads}
\author{P. W. Brouwer}
\address{Instituut-Lorentz, University of Leiden, P.O. Box 9506, 2300 RA
Leiden, The Netherlands}
\maketitle

\begin{abstract}
We consider the problem of the statistics of the scattering matrix $S$ of a
chaotic cavity (quantum dot), which is coupled to the outside world by
non-ideal leads containing $N$ scattering channels. The Hamiltonian $H$ of the
quantum dot is assumed to be an $M \times M$ hermitian matrix with probability
distribution $P(H) \propto \det[\lambda^2 + (H - \varepsilon)^2]^{-(\beta M + 2
- \beta)/2}$, where $\lambda$ and $\varepsilon$ are arbitrary coefficients and
$\beta = 1,2,4$ depending on the presence or absence of time-reversal and
spin-rotation symmetry. We show that this ``Lorentzian ensemble'' agrees with
microscopic theory for an ensemble of disordered metal particles in the limit
$M \rightarrow \infty$, and that for any $M \ge N$ it implies $P(S) \propto
|\det(1 - \bar S^{\dagger} S)|^{-(\beta M + 2 - \beta)}$, where $\bar S$ is the
ensemble average of $S$. This ``Poisson kernel'' generalizes Dyson's circular
ensemble to the case $\bar S \neq 0$ and was previously obtained from a maximum
entropy approach. The present work gives a microscopic justification for the
case that the chaotic motion in the quantum dot is due to impurity scattering.
\bigskip\\
\end{abstract}

\section{Introduction}

Recent experiments\cite{Marcus,Prober,Chang,Bird} on conductance fluctuations
and weak-localization effects in quantum dots have stimulated theoretical
work\cite{JSA,BJS,PEI,BarangerM,JPB,Pluhar,BrouwerB,Mucciolo} on phase-coherent
conduction through cavities in which the classical electron motion can be
regarded as chaotic. If the capacitance of the quantum dot is large enough, a
description in terms of non-interacting electrons is appropriate (otherwise the
Coulomb blockade becomes important\cite{JSA,PEI}).

For an isolated chaotic cavity, it has been conjectured and confirmed by many
examples that the statistics of the Hamiltonian $H$ agrees with that of the
Gaussian ensemble of random-matrix theory.\cite{Bohigas,Berry}
If the chaotic behavior is caused by impurity scattering, the agreement has
been established by microscopic theory: Both the Gaussian ensemble and the
ensemble of Hamiltonians with randomly placed impurities are equivalent to a
certain non-linear $\sigma$-model.\cite{Efetov,VWZ} Transport properties can be
computed by coupling $M$ eigenstates of $H$ to $N$ scattering
channels.\cite{VWZ,IWZ,Altland,LewenkopfWeidenmueller} Since $N \ll M$ this
construction introduces a great number of coupling parameters, whereas only a
few independent parameters determine the statistics of the scattering matrix
$S$ of the system.\cite{VWZ}

For transport properties at zero temperature and infinitesimal applied voltage,
one only needs to know $S$ at the Fermi energy $E_F$, and an approach which
starts directly from the ensemble of scattering matrices at a given energy is
favorable. Following up on earlier work on chaotic scattering in
billiards,\cite{BlumelSmilansky} two recent papers\cite{BarangerM,JPB} have
studied the transport properties of a quantum dot under the assumption that $S$
is distributed according to Dyson's circular ensemble.\cite{Dyson,Mehta} In
Refs.\ \ref{BarangerM} and \ref{JPB} the coupling of the quantum dot to the
external reservoirs was assumed to occur via ballistic point contacts (or
``ideal leads''). The extension to coupling via tunnel barriers (non-ideal
leads) was considered in Ref.\ \ref{BrouwerB}. In all cases complete agreement
was obtained with results which were obtained from the Hamiltonian
approach.\cite{PEI,IWZ,Altland} This agreement calls for a general
demonstration of the equivalence of the scattering matrix and the Hamiltonian
approach, for arbitrary coupling of the quantum dot to the external reservoirs.
It is the purpose of this paper to provide such a demonstration. A proof of the
equivalence of the Gaussian and circular ensembles has been published by
Lewenkopf and Weidenm\"uller,\cite{LewenkopfWeidenmueller} for the special case
of ideal leads. The present proof applies to non-ideal leads as well, and
corrects a subtle flaw in the proof of Ref.\ \ref{LewenkopfWeidenmueller} for
the ideal case.

The circular ensemble of scattering matrices is characterized by a probability
distribution $P(S)$ which is constant, that is to say, each unitary matrix $S$
is equally probable. As a consequence, the ensemble average $\bar S$ is zero.
This is appropriate for ideal leads. A generalization of the circular ensemble
which allows for non-zero $\bar S$ (and can therefore be applied to non-ideal
leads) has been derived by Mello, Pereyra, and
Seligman,\cite{MelloPereyraSeligman,MelloLesHouches} using a maximum entropy
principle. The distribution function in this generalized circular ensemble is
known in the mathematical literature\cite{Hua} as the Poisson kernel,
\begin{equation}
  P(S) \propto \left|\det(1 - \bar S^\dagger S)\right|^{-(\beta N + 2 -
\beta)}. \label{mainres}
\end{equation}
Here $\beta \in \{1,2,4\}$ is the symmetry index of the ensemble of scattering
matrices: $\beta = 1$ or $2$ in the absence or presence of a
time-reversal-symmetry breaking magnetic field; $\beta = 4$ in zero magnetic
field with strong spin-orbit scattering. (In Refs.\ \ref{MelloPereyraSeligman}
and \ref{MelloLesHouches} only the case $\beta = 1$ was considered.) One
verifies that $P(S) = \mbox{constant}$ for $\bar S = 0$. Eq.\ (\ref{mainres})
was first recognized as a possible generalization of the circular ensemble by
Krieger,\cite{Krieger} for the special case that $\bar S$ is proportional to
the unit matrix.

In this paper we present a microscopic justification of the Poisson kernel, by
deriving it from an ensemble of random Hamiltonians which is equivalent to an
ensemble of disordered metal grains. For the Hamiltonian ensemble we can use
the Gaussian ensemble, or any other ensemble to which it is equivalent in the
limit $M \rightarrow \infty$.\cite{WeidenmuellerPreprint} (The microscopic
justification of the Gaussian ensemble only holds for $M \rightarrow \infty$.)
For technical reasons, we use a Lorentzian distribution for the Hamiltonian
ensemble, which in the limit $M \rightarrow \infty$ can be shown to be
equivalent to the usual Gaussian distribution.
The technical advantage of the Lorentzian ensemble over the Gaussian ensemble
is that the equivalence to the Poisson kernel holds for arbitrary $M \ge N$,
and does not require taking the limit $M \rightarrow \infty$.

The outline of this paper is as follows: In Sec.\ \ref{sec3} the usual
Hamiltonian approach is summarized, following Ref.\ \ref{VWZ}. In Sec.\
\ref{sec2}, the Lorentzian ensemble is introduced. The eigenvalue and
eigenvector statistics of the Lorentzian ensemble are shown to agree with the
Gaussian ensemble in the limit $M \rightarrow \infty$. In Sec.\ \ref{sec4} we
then compute the entire distribution function $P(S)$ of the scattering matrix
from the Lorentzian ensemble of Hamiltonians, and show that it agrees with the
Poisson kernel (\ref{mainres}) for arbitrary $M \ge N$. In Sec.\ \ref{sec5} the
Poisson kernel is shown to describe a quantum dot which is coupled to the
reservoirs by means of tunnel barriers. We conclude in Sec.\ \ref{sec6}.

\section{Hamiltonian approach} \label{sec3}

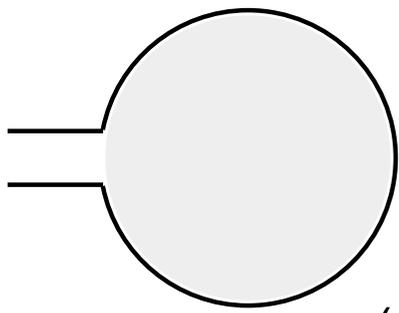
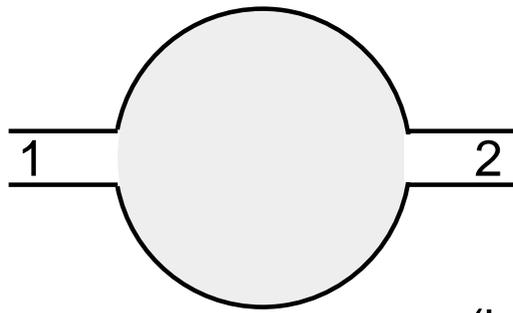
\begin{figure}
%\epsfysize=7cm
%\epsffile{cavfig8.eps}

\caption{Schematic drawing of a disordered cavity (grey) attached to a lead.
There are $N$ scattering channels in the lead, which are coupled to $M$ bound
levels in the cavity. In (a) only one lead is drawn. A system with more leads
(b) is described by combining them formally into one lead.\label{fig1}}
\end{figure}

The Hamiltonian approach\cite{VWZ,LewenkopfWeidenmueller,MahauxWeidenmueller}
starts with a formal division of the system into two parts, the leads and the
cavity (see Fig.\ \ref{fig1}a). The Hamiltonian of the total system is
represented in the following way: Let the set $\{|a\rangle\}$ represent a basis
of scattering states in the lead at the Fermi energy $E_F$ ($a = 1, \ldots,
N$), with $N$ the number of propagating modes at $E_F$. The set of bound states
in the cavity is denoted by $\{|\mu\rangle\}$ ($\mu = 1, \ldots, M$). We assume
$M \ge N$. The Hamiltonian $\cal H$ is then given by\cite{VWZ}
\begin{equation}
  {\cal H} = \sum_{a} |a\rangle E_F \langle a| + \sum_{\mu,\nu} |\mu\rangle
H_{\mu\nu}\langle\nu| + \sum_{\mu,a} \left( |\mu\rangle W_{\mu a} \langle a | +
|a\rangle W_{\mu a}^{*} \langle \mu | \right). \label{HamHam}
\end{equation}
The matrix elements $H_{\mu\nu}$ form a hermitian $M \times M$ matrix $H$, with
real ($\beta = 1$), complex ($\beta = 2$), or real quaternion ($\beta = 4$)
elements. The coupling constants $W_{\mu a}$ form a real (complex, real
quaternion) $M \times N$ matrix $W$. The $N \times N$ scattering matrix
$S(E_F)$ associated with this Hamiltonian is given by
\begin{equation}
  S(E_F) = 1 - 2 \pi i W^{\dagger} (E_F - H + i \pi W W^{\dagger})^{-1} W.
\label{SHeq}
\end{equation}
For $\beta = 1,2,4$ the matrix $S$ is respectively unitary symmetric, unitary,
and unitary self-dual.

Usually one assumes that $H$ is distributed according to the Gaussian ensemble,
\begin{equation}
  P(H) = {1 \over V} \exp\left(-\case{1}{4}\beta M \lambda^{-2} \mbox{tr}\,
H^2\right), \label{GaussEns}
\end{equation}
with $V$ a normalization constant and $\lambda$ an arbitrary coefficient which
determines the density of states at $E_F$. The coupling matrix $W$ is fixed.
Notice that $P(H)$ is invariant under transformations $H \rightarrow U H
U^{\dagger}$ where $U$ is orthogonal ($\beta = 1$), unitary ($\beta = 2$), or
symplectic ($\beta = 4$). This implies that $P(S)$ is invariant under
transformations $W \rightarrow U W$, so that it can only depend on the
invariant $W^{\dagger} W$. The ensemble-averaged scattering matrix $\bar S$ can
be calculated analytically in the limit $M \rightarrow \infty$, at fixed $N$,
$E_F$, and fixed mean level spacing $\Delta$. The result is\cite{VWZ}
\begin{equation}
  \bar S = {1 - \pi W^{\dagger} W/\lambda \over 1 + \pi W^{\dagger} W/\lambda}.
\label{SbarSupSym}
\end{equation}

It is possible to extend the Hamiltonian (\ref{HamHam}) to include a
``background'' scattering matrix $S_0$ which does not couple to the
cavity.\cite{NishiokaWeidenmueller} The matrix $S_0$ is symmetric for $\beta =
1$ and can be decomposed as $S_0 = O e^{2 i \Phi} O^T$, where the matrix $O$ is
orthogonal and $\Phi$ is real and diagonal. In the limit $M \rightarrow
\infty$, the average scattering matrix $\bar S$ is now given
by\cite{NishiokaWeidenmueller}
\begin{equation}
  \bar S = O e^{i \Phi} {1 - \pi W^{\dagger} W/\lambda \over 1 + \pi
W^{\dagger} W/\lambda } e^{i \Phi} O^T. \label{SbarSupSymExt}
\end{equation}

Lewenkopf and Weidenm\"uller\cite{LewenkopfWeidenmueller} used this extended
version of the theory to relate the Gaussian and circular ensembles, for $\beta
= 1$ and $\bar S = 0$. Their argument is based on the assumption that Eq.\
(\ref{SbarSupSymExt}) can be inverted, to yield $W^{\dagger} W$ and $S_0$ as a
function of $\bar S$. Then $P(S) = P_{\bar S}(S)$ is fully determined by $\bar
S$ (and does not require separate knowledge of $W^{\dagger} W$ and $S_0$).
Under the transformation $S \rightarrow U S U^T$ (with $U$ an arbitrary unitary
matrix), $\bar S$ is mapped to $U \bar S U^T$, which implies
\begin{equation}
  P_{\bar S}(S) = P_{U \bar S U^T} (U S U^T).
\end{equation}
For $\bar S = 0$ one finds that $P(S)$ is invariant under transformations $S
\rightarrow U S U^T$, so that $P(S)$ must be constant (circular ensemble).
There is, however, a weak spot in this argument: Equation (\ref{SbarSupSymExt})
can {\em not} be inverted for the crucial case $\bar S = 0$. It is only
possible to determine $W^{\dagger} W$, not $S_0$. This is a serious objection,
since $S_0$ is not invariant under the transformation $S \rightarrow U S U^T$,
and one can not conclude that $P(S) = \mbox{constant}$ for $\bar S = 0$. We
have not succeeded in repairing the proof of Ref.\ \ref{LewenkopfWeidenmueller}
for $\bar S = 0$, and instead present in the following sections a different
proof (which moreover can be extended to non-zero $\bar S$).

A situation in which the cavity is coupled to $n$ reservoirs by $n$ leads,
having $N_j$ scattering channels ($j = 1,\ldots,n$) each, can be described in
the framework presented above by combining the $n$ leads formally into a single
lead with $N  = \sum_{j=1}^{n} N_j$ scattering channels. Scattering matrix
elements between channels in the same lead correspond to reflection from the
cavity, elements between channels in different leads correspond to
transmission. In this notation, the Landauer formula for the conductance $G$ of
a cavity with two leads (Fig.\ \ref{fig1}b) takes the form
\begin{equation}
  G = {2e^2 \over h} \sum_{i = 1}^{N_1} \sum_{j=N_1+1}^{N_1+N_2} |S_{ij}|^2.
\end{equation}

\section{Lorentzian ensemble} \label{sec2} \label{SEC2}

For technical reasons we wish to replace the Gaussian distribution
(\ref{GaussEns}) of the Hamiltonians by a Lorentzian distribution,
\begin{equation}
  P(H) = {1 \over V} \lambda^{M(\beta M + 2 - \beta)/2} \det \left( \lambda^2 +
(H-\varepsilon)^2 \right)^{-(\beta M + 2 - \beta)/2}, \label{LorEns}
\end{equation}
where $\lambda$ and $\varepsilon$ are parameters describing the width and
center of the distribution, and $V$ is a normalization constant independent of
$\lambda$ and $\varepsilon$. The symmetry parameter $\beta \in \{1,2,4\}$
indicates whether the matrix elements of $H$ are real [$\beta = 1$, Lorentzian
Orthogonal Ensemble ($\Lambda$OE)], complex [$\beta = 2$, Lorentzian Unitary
Ensemble ($\Lambda$UE)], or real quaternion [$\beta = 4$, Lorentzian Symplectic
Ensemble ($\Lambda$SE)]. (We abbreviate ``Lorentzian'' by a capital lambda,
because the letter $L$ is commonly used to denote the Laguerre ensemble.)

The replacement of (\ref{GaussEns}) by (\ref{LorEns}) is allowed because the
eigenvector and eigenvalue distributions of the Gaussian and the Lorentzian
ensemble are equal on a fixed energy scale, in the limit $M \rightarrow \infty$
at a fixed mean level spacing $\Delta$. The equivalence of the eigenvector
distributions is obvious: The distribution of $H$ depends solely on the
eigenvalues for both the Lorentzian and the Gaussian ensemble, so that the
eigenvector distribution is uniform for both ensembles. In order to prove the
equivalence of the distribution of the eigenvalues $E_1, E_2, \ldots, E_M$
(energy levels), we compare the $n$-level cluster functions $T_n(E_1, E_2,
\ldots,E_n)$ for both ensembles. The general definition of the $T_n$'s is given
in Ref.\ \ref{Mehta}. The first two $T_n$'s are defined by
\begin{mathletters}
\begin{eqnarray}
  T_1(E) &=& \langle \sum_{i=1}^{M} \delta(E - E_i) \rangle \\
  T_2(E_1,E_2) &=& \langle \sum_{i,j=1}^{M} \delta(E_1 - E_i) \delta(E_2 - E_j)
\rangle - \langle \sum_{i=1}^{M} \delta(E_1 - E_i) \rangle \langle
\sum_{j=1}^{M} \delta(E_2 - E_j) \rangle.
\end{eqnarray}
\end{mathletters}%
The brackets $\langle \ldots \rangle$ denote an average over the ensemble. The
cluster functions in the Gaussian ensemble are known for arbitrary
$n$,\cite{Mehta} for the Lorentzian ensemble we compute them below.

{}From Eq.\ (\ref{LorEns}) one obtains the joint probability distribution
function of the eigenvalues,
\begin{equation}
  P(\{E_j\}) = {1 \over V} \lambda^{M(\beta M + 2 - \beta)/2}\prod_{i < j} |E_i
- E_j|^{\beta} \prod_i \left({\lambda^2 + (E_i -
\varepsilon)^2)}\right)^{-(\beta M + 2 - \beta)/2}. \label{LorEnsE}
\end{equation}
We first consider the case $\lambda = 1$, $\varepsilon = 0$. We make the
transformation
\begin{equation}
  S = {1 + i H \over 1 - i H}.
\end{equation}
The eigenvalues $e^{i \phi_j}$ of the unitary matrix $S$ are related to the
energy levels $E_j$ by
\begin{equation}
  e^{i \phi_j} = {1 + i E_j \over 1 - i E_j} \ \Longleftrightarrow \ \phi_j = 2
\arctan{E_j}.
  \label{eigenphaseigenval1}
\end{equation}
The probability distribution of the eigenphases follows from Eqs.\
(\ref{LorEnsE}) and (\ref{eigenphaseigenval1}),
\begin{equation}
  P(\{\phi_j\}) = {1 \over V} 2^{-M(\beta M + 2 - \beta)/2} \prod_{i < j} |
e^{i \phi_i} - e^{i \phi_j} |^{\beta}.
\end{equation}
This is precisely the distribution of the eigenphases in the circular ensemble.
The cluster functions in the circular ensemble are known.\cite{Dyson,Mehta}
The $n$-level cluster functions $T^{\Lambda}_n$ in the Lorentzian ensemble are
thus related to the $n$-level cluster functions $T^{C}_n$ in the circular
ensemble by
\begin{equation}
  T^{\Lambda}_n(E_1,\ldots,E_n) = T^{C}_n(2 \arctan E_1,\ldots,2 \arctan E_n)
\prod_{j = 1}^{n} {2 \over 1 + E_j^2}. \label{kappaE1}
\end{equation}
For $n=1$ one finds the level density
\begin{equation}
  \rho(E) = {M \over \pi(1 + E^2)}, \label{rhoE1}
\end{equation}
independent of $\beta$. For $n=2$ one finds the pair-correlation function
\begin{equation}
  T^{\Lambda}_2(E_1, E_2) = {4 \, \sin^2 ( M \arctan{E_1}- M\arctan{E_2} )
\over (1 + E_1^2)(1 + E_2^2)\, \sin^2(\arctan{E_1}- \arctan{E_2})}.
\label{PairLor}
\end{equation}
Eq.\ (\ref{PairLor}) holds for $\beta = 2$. The expressions for $\beta = 1,4$
are more complicated.

The $n$-level cluster functions for arbitrary $\lambda$ and $\varepsilon$ can
be found after a proper rescaling of the energies. Eq.\ (\ref{kappaE1})
generalizes to
\begin{eqnarray}
  T^{\Lambda}_n(E_1,\ldots,E_n) &=& T^{C}_n\left(2 \arctan\frac{E_1 -
\varepsilon}{\lambda},\ldots,2\arctan \frac{E_n - \varepsilon}{\lambda} \right)
\prod_{j = 1}^{n} {2 \lambda \over \lambda^2 + (E_j - \varepsilon)^2}.
\label{LorCluFunc}
\end{eqnarray}

The large-$M$ limit of the $T_n$'s is defined as
\begin{equation}
  Y_n (\xi_1,\ldots,\xi_n) = \lim_{M \rightarrow \infty} \Delta^n T_n(\xi_1
\Delta, \ldots, \xi_n \Delta).
\end{equation}
For both the Gaussian and the Lorentzian ensembles, the mean level spacing
$\Delta$ at the center of the spectrum in the limit $M \rightarrow \infty$ is
given by $\Delta = \lambda \pi / M$. Therefore, the relevant limit $M
\rightarrow \infty$ at fixed level spacing is given by $M \rightarrow \infty$,
$\lambda \rightarrow \infty$, $\Delta = \lambda \pi / M$ fixed for both
ensembles. Equation (\ref{LorCluFunc}) allows us to relate the $Y_n$'s in the
Lorentzian and circular ensembles,
\begin{eqnarray}
  Y^{\Lambda}_n(\xi_1,\ldots,\xi_n) &=& \lim_{M \rightarrow \infty} \left( {2
\pi / M} \right)^n T_n^{C}(2 \arctan(\pi\xi_1/M), \ldots, 2
\arctan(\pi\xi_n/M)) \nonumber \\ &=& \lim_{M \rightarrow \infty} \left( {2 \pi
/ M} \right)^n T_n^{C}(2 \pi \xi_1/M, \ldots, 2 \pi \xi_n/M) \nonumber \\ &=&
Y_n^{C}(\xi_1, \ldots, \xi_n). \label{ClusterLC}
\end{eqnarray}
It is known that the cluster functions $Y_n^{C}$ in the circular ensemble are
equal to the cluster functions $Y_n^{G}$ in the Gaussian ensemble.\cite{Mehta}
Equation (\ref{ClusterLC}) therefore shows that the Lorentzian and the Gaussian
ensembles have the same cluster functions in the large-$M$ limit.

The technical reason for working with the Lorentzian ensemble instead of with
the Gaussian ensemble is that the Lorentzian ensemble has two properties which
make it particularly easy to compute the distribution of the scattering matrix.
The two properties are:

{}~\\
{\bf Property 1:} If $H$ is distributed according to a Lorentzian ensemble with
width $\lambda$ and center $\varepsilon$, then $H^{-1}$ is again distributed
according to a Lorentzian ensemble, with width $\tilde \lambda = {\lambda /
(\lambda^2 + \varepsilon^2)}$ and center $\tilde \varepsilon = {\varepsilon /
(\lambda^2 + \varepsilon^2)} $.

{}~\\
{\bf Property 2:}  If the $M \times M$ matrix $H$ is distributed according to a
Lorentzian ensemble, then every $N \times N$ submatrix of $H$ obtained by
omitting $M - N$ rows and the corresponding columns is again distributed
according to a Lorentzian ensemble, with the same width and center.

{}~\\
The proofs of both properties are essentially contained in Ref.\ \ref{Hua}. In
order to make this paper self-contained, we briefly give the proofs in the
appendix.

\section{Scattering matrix distribution for the Lorentzian ensemble}
\label{sec4}

The general relation between the Hamiltonian $H$ and the scattering matrix $S$
is given by Eq.\ (\ref{SHeq}). After some matrix manipulations, it can be
written as
\begin{equation}
  S = \left(1 + i \pi W^{\dagger} (H-E_F)^{-1} W\right)
      \left(1 - i \pi W^{\dagger} (H-E_F)^{-1} W\right)^{-1}.
\end{equation}
We can write the coupling matrix $W$ as
\begin{equation}
  W = U Q \tilde{W}, \label{Qdef}
\end{equation}
where $U$ is an $M \times M$ orthogonal ($\beta = 1$), unitary ($\beta = 2$),
or symplectic ($\beta = 4$) matrix, $\tilde{W}$ is an $N \times N$ matrix, and
$Q$ is an $M \times N$ matrix with all elements zero except $Q_{nn} = 1$, $1
\le n \le N$. Substitution into Eq.\ (\ref{SHeq}) gives
\begin{equation}
  S = \left(1 + i \pi \tilde{W}^{\dagger} \tilde H \tilde{W}\right)
  \left(1 - i \pi \tilde{W}^{\dagger} \tilde H \tilde{W}\right)^{-1},
  \label{SNN}
\end{equation}
where we have defined $\tilde H \equiv Q^T U^{\dagger} (H-E_F)^{-1} U Q$.

We assume that $H$ is a member of the Lorentzian ensemble, with width $\lambda$
and center $0$. Then the matrix $H - E_F$ is also a member of the Lorentzian
ensemble, with width $\lambda$ and center $E_F$. Property 1 implies that $(H -
E_F)^{-1}$ is distributed according to a Lorentzian ensemble with width $\tilde
\lambda = \lambda/(\lambda^2 + E_F^2)$ and center $\tilde \varepsilon =
E_F/(\lambda^2 + E_F^2)$. Orthogonal (unitary, symplectic) invariance of the
Lorentzian ensemble implies that $U^{\dagger} (H-E_F)^{-1} U$ has the same
distribution as $(H-E_F)^{-1}$. Using property 2 we then find that $\tilde H$
[being an $N \times N$ submatrix of $U^{\dagger}(H - E_F)^{-1}U$] is
distributed according to the same Lorentzian ensemble (width $\tilde \lambda$
and center $\tilde \varepsilon$).

We now compute the distribution of the scattering matrix, first for a special
coupling, then for the general case.

\subsection{Special coupling matrix}

First we will consider the special case that
\begin{equation}
  \tilde W = \pi^{-1/2} \delta_{nm} \label{Wspecial}
\end{equation}
is proportional to the unit matrix. The relation (\ref{SNN}) between the $S$
and $\tilde H$ is then
\begin{equation}
  S = {1 + i \tilde H \over 1 - i \tilde H}.
\end{equation}
Thus the eigenvalues $\tilde E_j$ of $\tilde H$ and $e^{i \phi_j}$ of $S$ are
related via
\begin{equation}
  e^{i \phi_j} = {1 + i \tilde E_j \over 1 - i \tilde E_j} \
\Longleftrightarrow \ \phi_j = 2 \arctan{\tilde E_j}. \label{eigenphaseigenval}
\end{equation}

Since transformations $\tilde H \rightarrow U \tilde H U^{\dagger}$ (with
arbitrary orthogonal, unitary, or symplectic $N \times N$ matrix $U$) leave
$P(\tilde H)$ invariant, $P(S)$ is also invariant under $S \rightarrow U S
U^{\dagger}$. So $P(S)$ can only depend on the eigenvalues $e^{i \phi_j}$ of
$S$. The distribution of the $\tilde E$'s is [cf.\ Eq.\ (\ref{LorEnsE})]
\begin{equation}
  P(\{\tilde E_j\}) = {1 \over V} \lambda^{N(\beta N + 2 - \beta)/2} \prod_{j <
k} |\tilde E_j - \tilde E_k|^{\beta} \prod_{j} \left({\tilde \lambda^2 +
(\tilde E_j - \tilde \varepsilon)^2}\right)^{-(\beta N + 2 - \beta)/2}.
\label{eigenvaltildeH}
\end{equation}
{}From Eqs.\ (\ref{eigenphaseigenval}) and (\ref{eigenvaltildeH}) we obtain the
probability distribution of the $\phi$'s,
\begin{mathletters}
  \label{KE}
\begin{eqnarray}
  P(\{\phi_j\}) &=& {1 \over V} \left({1 - \sigma \sigma^{*} \over
2}\right)^{N(\beta N + 2 - \beta)/2} \prod_{j < k} |e^{i \phi_j} - e^{i
\phi_k}|^{\beta} \ \prod_{j} \left|1 - \sigma^{*} e^{i
\phi_{j}}\right|^{-(\beta N + 2 - \beta)}, \\
  \sigma &=& {1 - \tilde \lambda - i \tilde \varepsilon \over 1 + \tilde
\lambda + i \tilde \varepsilon} = {\lambda^2 + E_F^2 - \lambda - i E_F \over
\lambda^2 + E_F^2 + \lambda + i E_F}.
\end{eqnarray}
\end{mathletters}%
Eq.\ (\ref{KE}) implies that $P(S)$ has the form of a Poisson kernel,
\begin{equation}
  P(S) = {\det(1 - \bar S \bar S^{\dagger})^{(\beta N + 2 - \beta)/2} \over
2^{N(\beta N + 2 - \beta)/2} V} \left|\det(1 - \bar S^{\dagger}
S)\right|^{-(\beta N + 2 - \beta)}, \label{P0distr}
\end{equation}
the average scattering matrix $\bar S$ being given by
\begin{equation}
  \bar S_{nm} = \sigma \delta_{nm}.
\end{equation}

\subsection{Arbitrary coupling matrix}

Now we turn to the case of arbitrary coupling matrix $\tilde W$. We denote the
scattering matrix at coupling $\tilde W$ by $S$, and denote the scattering
matrix at the special coupling (\ref{Wspecial}) by $S_0$. The relation between
$S$ and $S_0$ is
\begin{equation}
  S  = r + t' S_0 (1 - r' S_0)^{-1} t \label{SfromS0} \Longleftrightarrow S_0 =
(t')^{-1} (S - r) (1 - r^{\dagger} S)^{-1} t^{\dagger}, \label{S0fromS}
\end{equation}
where we abbreviated
\begin{mathletters}
  \label{StransformW}
\begin{eqnarray}
  r  &=& (1 - \pi \tilde W^{\dagger} \tilde W)(1 + \pi \tilde W^{\dagger}
\tilde W)^{-1}, \\
  r' &=& -\tilde W (1 - \pi \tilde W^{\dagger} \tilde W)(1 + \pi \tilde
W^{\dagger} \tilde W)^{-1} \tilde W^{-1}, \\
  t  &=& 2 \pi^{1/2} \tilde W (1 + \pi \tilde W^{\dagger} \tilde W)^{-1}, \\
  t' &=& 2 \pi^{1/2} (1 + \pi \tilde W^{\dagger} \tilde W)^{-1} \tilde
W^{\dagger}.
\end{eqnarray}
\end{mathletters}%
The symmetry of the coupling matrix $\tilde W$ is reflected in the symmetry of
the $2N \times 2N$ matrix
\begin{equation}
  S_1 = \left( \begin{array}{cc} r & t' \\ t & r' \end{array} \right),
\label{S1decomp}
\end{equation}
which is unitary symmetric ($\beta = 1$), unitary ($\beta = 2$) or unitary
self-dual ($\beta = 4$).

The probability distribution $P_0$ of $S_0$ is given by Eq.\ (\ref{P0distr}).
The distribution $P$ of $S$ follows from
\begin{mathletters} \label{Probrel}
\begin{equation}
  P(S) = P_0(S_0) {d S_0 \over d S},
\end{equation}
where the Jacobian $dS_0/dS$ is the ratio of infinitesimal volume elements
around $S_0$ and $S$. This Jacobian is known,\cite{Hua,FriedmanMello}
\begin{equation} \label{VolRel}
  {d S_0 \over d S} = \left(
  {\det(1 - r^{\dagger} r)
  \over
  |\det(1 - r^{\dagger} S)|^2} \right)^{(\beta N + 2 - \beta)/2}.
\end{equation}
\end{mathletters}%
After expressing $S_0$ in terms of $S$ by means of Eq.\ (\ref{S0fromS}), we
find that $P(S)$ is given by the same Poisson kernel as Eq.\ (\ref{P0distr}),
but with a different $\bar S$,
\begin{equation}
  \bar S = {1 - \pi (\tilde \lambda + i \tilde \varepsilon) W^{\dagger} W
\over
            1 + \pi (\tilde \lambda + i \tilde \varepsilon) W^{\dagger} W }.
  \label{SbarLorEnsArbW}
\end{equation}
In the limit $M \rightarrow \infty$ at fixed level spacing $\Delta = \lambda
\pi /M$, Eq.\ (\ref{SbarLorEnsArbW}) simplifies to
\begin{equation}
  \bar S = {\Delta M - \pi^2 W^{\dagger} W  \over
            \Delta M + \pi^2 W^{\dagger} W }.
  \label{SbarDeltaArbW}
\end{equation}

The extended version of the Hamiltonian approach which includes a background
scattering matrix $S_0$ can be mapped to the case without background scattering
matrix by a transformation $S \rightarrow S'= U S U^T$ ($\beta = 1$), $S
\rightarrow S' = U S V$ ($\beta = 2$), or $S \rightarrow S' = U S U^R$ ($\beta
= 4$), where $U$ and $V$ are unitary matrices.\cite{NishiokaWeidenmueller}
($U^T$ is the transposed of $U$, $U^R$ is the dual of $U$.) The Poisson kernel
is covariant under such transformations,\cite{MelloPereyraSeligman} i.e. it
maps to a Poisson kernel with $\bar S' = U \bar S U^T$ ($\beta = 1$), $\bar S'
= U \bar S V$ ($\beta = 2$), or $\bar S' = U \bar S U^R$ ($\beta = 4$). As a
consequence, the distribution of $S$ is given by the Poisson kernel for
arbitrary coupling matrix $W$ and background scattering matrix $S_0$. This
proves the general equivalence of the Poisson kernel and the Lorentzian
ensemble of Hamiltonians.

\section{Ideal versus non-ideal leads} \label{sec5}

The circular ensemble of scattering matrices is appropriate for a chaotic
cavity which is coupled to the leads by means of ballistic point contacts
(``ideal'' leads). In this section we will demonstrate that the generalized
circular ensemble described by the Poisson kernel is the appropriate ensemble
for a chaotic cavity which is coupled to the leads by means of tunnel barriers
(``non-ideal'' leads).

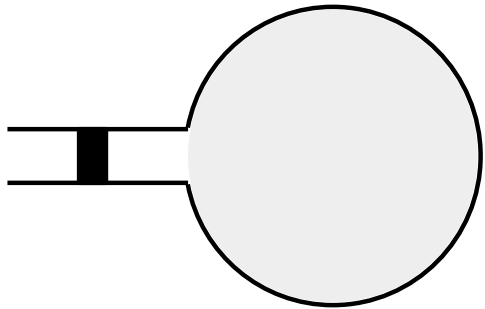
\begin{figure}
%\epsfysize=7cm
%\epsffile{cavfig2.eps}

\caption{Schematic drawing of the chaotic cavity and the non-ideal lead
containing a tunnel barrier.\label{fig2}}
\end{figure}

The system considered is shown schematically in Fig.\ \ref{fig2}. We assume
that the segment of the lead between the tunnel barrier and the cavity is long
enough, so that both the $N \times N$ scattering matrix $S_0$ of the cavity and
the $2 N \times 2 N$ scattering matrix $S_1$ of the tunnel barrier are
well-defined. The scattering matrix $S_0$ has probability distribution $P_0 =
\mbox{constant}$ of the circular ensemble, whereas the scattering matrix $S_1$
is kept fixed.

We decompose $S_1$ in terms of $N \times N$ reflection and transmission
matrices,
\begin{equation}
  S_1 = \left( \begin{array}{cc} r_1^{\vphantom{2}} & t_1' \\
t_1^{\vphantom{2}} & r_1' \end{array} \right).
\end{equation}
The $N \times N$ scattering matrix $S$ of the total system is related to $S_0$
and $S_1$ by
\begin{equation}
  S = r_1 + t_1' (1 - S_0 r_1')^{-1} S_0 t_1. \label{SfromS01}
\end{equation}
This relation has the same form as Eq.\ (\ref{SfromS0}). We can therefore
directly apply Eq. (\ref{Probrel}), which yields
\begin{equation}
  P(S) \propto |\det(1 - r_1^{\dagger} S)|^{-(\beta N + 2 - \beta)}.
\end{equation}
Hence $S$ is distributed according to a Poisson kernel, with $\bar S = r_1$.

\section{Conclusion} \label{sec6}

In conclusion we have established by explicit computation the equivalence for
$M \ge N$ of a generalized circular ensemble of scattering matrices (described
by a Poisson kernel) and an ensemble of $M \times M$ Hamiltonians with a
Lorentzian distribution. The Lorentzian and Gaussian distributions are
equivalent in the large-$M$ limit. Moreover, the Gaussian Hamiltonian ensemble
and the microscopic theory of a metal particle with randomly placed impurities
give rise to the same non-linear $\sigma$-model.\cite{Efetov,VWZ} Altogether,
this provides a microscopic justification of the Poisson kernel in the case
that the chaotic motion in the cavity is caused by impurity scattering. For the
case of a ballistic chaotic cavity, a microscopic justification is still
lacking.

The equivalence of the Poisson kernel and an arbitrary Hamiltonian ensemble can
be reformulated in terms of a central limit theorem: The distribution of a
submatrix of $H^{-1}$ of fixed size $N$ tends to a Lorentzian distribution when
$M \rightarrow \infty$, independent of the details of the distribution of $H$.
A central limit theorem of this kind for $N = 1$ has previously been formulated
and proved by Mello.\cite{MelloLesHouches}

This work was motivated by a series of lectures by P.\ A.\ Mello at the
summerschool in Les Houches on ``Mesoscopic Quantum Physics''. Discussions with
C.\ W.\ J.\ Beenakker, K. Frahm, P.\ A.\ Mello, and H.\ A.\ Weidenm\"uller are
gratefully acknowledged. This research was supported by the ``Stich\-ting voor
Fun\-da\-men\-teel On\-der\-zoek der Ma\-te\-rie'' (FOM) and by the
``Ne\-der\-land\-se or\-ga\-ni\-sa\-tie voor We\-ten\-schap\-pe\-lijk
On\-der\-zoek'' (NWO).

\appendix

\section*{Proof of properties 1 and 2 of Sec.\ \protect\ref{sec2}}

The two proofs given below are adapted from Ref.\ \ref{Hua}. The matrix $H$ and
its inverse $H^{-1}$ have the same eigenvectors, but reciprocal eigenvalues.
Therefore, property 1 of the Lorentzian ensemble is proved by showing that the
distribution of the eigenvalues of $H^{-1}$ is given by Eq.\ (\ref{LorEnsE}),
with the substitutions $\lambda \rightarrow \tilde \lambda$ and $\varepsilon
\rightarrow \tilde \varepsilon$. This is easily done,
\begin{eqnarray}
  P(\{E_j^{-1}\}) &=&
  {1 \over V} \lambda^{M(\beta M + 2 - \beta)/2} \prod_{i < j} |E_i -
E_j|^{\beta} \prod_i \left[ \left({\lambda^2 + (E_i -
\varepsilon)^2}\right)^{-(\beta M + 2 - \beta)/2} \left| {d E_i \over d
(E_i^{-1})} \right| \right] \nonumber \\ &=&
  {1 \over V} \lambda^{M(\beta M + 2 - \beta)/2} \prod_{i < j} \left|E_i E_j
(E_i^{-1} - E_j^{-1}) \right|^{\beta} \prod_i \left[ \left({\lambda^2 + (E_i -
\varepsilon)^2}\right)^{-(\beta M + 2 - \beta)/2} E_i^{2} \right] \nonumber \\
&=&
  {1 \over V} \lambda^{M(\beta M + 2 - \beta)/2} \prod_{i < j} \left|E_i^{-1} -
E_j^{-1}\right|^{\beta} \prod_i \left({\lambda^2 E_i^{-2} + (1 - \varepsilon
E_i^{-1})^2}\right)^{-(\beta M + 2 - \beta)/2} \nonumber \\ &=&
  {1 \over V} \tilde \lambda^{M(\beta M + 2 - \beta)/2}\prod_{i < j}
\left|E_i^{-1} - E_j^{-1}\right|^{\beta} \prod_i \left({\tilde \lambda^2 +
(E_i^{-1} - \tilde \varepsilon)^2}\right)^{-(\beta M + 2 - \beta)/2}.
\end{eqnarray}

In order to prove property 2, we may assume that after rescaling of $H$ we have
$\lambda = 1$, $\varepsilon = 0$. First consider $M = N - 1$. In this case, one
can write
\begin{equation}
  H = \left( \begin{array}{cc} G & Y \\ Y^{\dagger} & Z \end{array} \right),
\end{equation}
where $G$ is the $N \times N$ submatrix of $H$ whose distribution we want to
compute, $Y$ is a vector, with real ($\beta = 1$), complex ($\beta = 2$), or
real quaternion elements ($\beta = 4$), and $Z$ is a real number. For the
successive integrations over $Z$ and $Y$ we need two auxiliary results. First,
for real numbers $a$, $b$, $c$ such that $a > 0$ and $4 a c > b^2$, and for
real $m > 2$ we have
\begin{equation}
  \int_{-\infty}^{\infty} dx\, (a x^2 + bx + c)^{-m} = \left( {a c -
\case{1}{4}b^2} \right)^{-m + 1/2} a^{m-1} \pi^{1/2} {\Gamma(m - 1/2) /
\Gamma(m)}. \label{lemma1eq}
\end{equation}
Second, if $x$ is a $d$-dimensional vector with real components, and if $m >
(d+1)/2$, then
\begin{equation}
  \int_{-\infty}^{\infty} dx_1\ldots\int_{-\infty}^{\infty} dx_d\,
  {(1+x^2)^{-m}} =
  \pi^{d/2} {\Gamma(m - d/2) / \Gamma(m)}. \label{lemma2eq}
\end{equation}

Since $\det(1 + \tilde H)^2$ is a quadratic function of $Z$, the integral over
$Z$ can now be carried out using Eq.\ (\ref{lemma1eq}). The result is:
\begin{eqnarray}
\int dZ P(\tilde H) &=&
  {\pi^{1/2} \Gamma\left(\case{1}{2}(\beta M + 1 - \beta)\right) \over V
\Gamma\left(\case{1}{2}(\beta M + 2 - \beta)\right)} \det(1 + G^2)^{(-\beta M -
2 + \beta)/2} \times \nonumber \\ && \ \left( 1 + Y^{\dagger} (1 + G^2)^{-1} Y
\right)^{-\beta M - 1 + \beta}.
\end{eqnarray}
Next, we integrate over $Y$. We may choose the basis for the $Y$-vectors so
that $1 + G^2$ is diagonal, with diagonal elements $1 + G_i^2$. After rescaling
of the $Y$-vectors to $Y_i' = Y_i (1 + G_i^2)$ one obtains an integral similar
to Eq.\ (\ref{lemma2eq}), with $d = \beta (M-1)$. The final result is
\begin{eqnarray}
  P(G) =
{\pi^{(\beta M - \beta + 1)/2}\, \Gamma\left(\case{1}{2}(\beta M + 1 -
\beta)\right) \over V \Gamma\left(\beta M  + 1- \beta\right)} \det(1 +
G^2)^{(-\beta (M-1) -2 + \beta)/2}. \label{finproof2}
\end{eqnarray}
Property 2 now follows by induction. Notice that Eq.\ (\ref{finproof2}) allows
us to determine the normalization constant $V$,
\begin{equation}
  V = \pi^{(\beta M- \beta + 2)M/4}\prod_{j = 1}^{M}
{\Gamma\left(\case{1}{2}(\beta j + 1 - \beta)\right) \over \Gamma\left(\beta j
+ 1 - \beta \right)}.
\end{equation}

\end{document}